\documentclass[%
 reprint,
 amsmath,amssymb,
 aps,
]{revtex4-2}
\usepackage{geometry}
\usepackage{subcaption}
\usepackage{graphicx}
\usepackage{dcolumn}
\usepackage{bm}
\usepackage{booktabs}
\usepackage{caption}
\usepackage{lipsum}
\usepackage{tabularx}
\usepackage{adjustbox}
\usepackage{rotating}
\usepackage{bm}
\usepackage[utf8]{inputenc}
\usepackage{amsmath,amssymb}
\usepackage{tikz}
\usetikzlibrary{arrows.meta}
\usepackage{hyperref}   

\begin{document}
\preprint{APS/123-QED}
\title{$B_{sJ}(6114)$ as a Higher-Partial-Wave $\bar{K}^{*}B^{(*)}$ Molecular Candidate}
\thanks{}
\author{Shiquan Zhang}
\author{Dan Jiang}
\author{Yin Huang\footnote{corresponding author}}
\email{huangy2019@swjtu.edu.cn}
\affiliation{School of Physical Science and Technology, Southwest Jiaotong University, Chengdu 610031, China}

\begin{abstract}
In 2021, the LHCb Collaboration observed the $B_{sJ}(6114)$ state in the $B^\pm K^\mp$ invariant mass spectrum. It is currently the highest-mass bottom-strange state listed by the Particle Data Group (PDG). However, its spin-parity quantum numbers have not yet been experimentally determined, and its internal structure remains under debate. Although several quark-model interpretations have been proposed, including the $1^3D_1$ state and a member of the $1D(2^-,3^-)$ doublet, the possibility of a hadronic molecular structure for this state has not yet been systematically investigated.
We investigate the possibility of interpreting \(B_{sJ}(6114)\) as a \(B^{(*)}\bar{K}^{*}\) molecular state within the one-boson-exchange model. By solving the Schrödinger equation using the Gaussian expansion method, we study the bound-state properties with different spin-parity assignments, taking into account partial-wave mixing and coupled-channel effects. We further calculate the strong decay widths into the allowed \(B^{(*)}\bar{K}\) channels and compare them with the experimental data. Our results show that the mass spectrum alone cannot uniquely determine the spin-parity of \(B_{sJ}(6114)\). However, the decay-width analysis disfavors the \(J^P=0^+\), \(0^-\), \(1^+\), \(2^+\), and \(3^-\) assignments, while the \(J^P=1^-\), \(2^-\), and \(3^+\) configurations remain compatible with the experimental constraints. Among these possibilities, the \(J^P=1^-\) assignment provides the best agreement with the measured decay width. These findings suggest that \(B_{sJ}(6114)\) may be a higher-partial-wave \(B^{(*)}\bar{K}^{*}\) molecular state.
\end{abstract}
\date{\today}
\maketitle

\section{\label{sec:level1}introduction}
Over the past three decades, with the rapid development of experimental techniques, 
a large number of hadronic states have been observed~\cite{ParticleDataGroup:2024cfk}, 
greatly enriching our understanding of the hadron spectrum. However, despite the 
successful exploration of the charmed-strange meson sector, where 11 states have been 
experimentally established~\cite{ParticleDataGroup:2024cfk}, the corresponding bottom-strange partners expected from 
heavy-quark flavor symmetry remain largely unexplored, with only a limited number of 
states having been identified experimentally~\cite{ParticleDataGroup:2024cfk}.  A prominent example is the $D_{s0}^{\ast}(2317)$ and $D_{s1}(2460)$ states. These two 
states, which lie below the conventional quark-model expectations~\cite{Godfrey:1985xj}, are widely interpreted 
as having significant $DK$ and $D^{\ast}K$ molecular components~\cite{BaBar:2003oey,CLEO:2003ggt,
Belle:2003kup,Belle:2003guh,BaBar:2004yux,Meng:2022ozq,Xie:2010zza,Guo:2006fu,
Guo:2006rp,Gamermann:2006nm,Zhu:2019vnr,Mohler:2013rwa,Altenbuchinger:2013vwa}. 
Based on the heavy-quark symmetry~\cite{Isgur:1991wq}, their bottom counterparts, $B_{s0}^{\ast}$ and 
$B_{s1}^{\ast}$, are expected to possess analogous $K\bar{B}$ and $K\bar{B}^{\ast}$ 
molecular structures~\cite{Guo:2006fu,Guo:2006rp,Wang:2019ehs,Cheng:2014bca,
Colangelo:2012xi,Kolomeitsev:2003ac,Yang:2022vdb,Lang:2015hza,Albaladejo:2016ztm,
Altenbuchinger:2013vwa,Vishwakarma:2022sly}. Nevertheless, these predicted bottom-strange states have not yet been experimentally established~\cite{D0:2016mwd,ATLAS:2018udc,CMS:2017hfy,LHCb:2015aaf,LHCb:2016dxl,CDF:2017dwr}, making the search for these molecular states and other bottom-strange states a subject of intensive research in heavy-flavor hadron spectroscopy.

In 2021, the LHCb Collaboration reported the observation of the $B_{sJ}(6114)$ state~\cite{LHCb:2020pet}, 
which is currently the highest-mass bottom-strange state listed by the Particle Data Group (PDG)~\cite{ParticleDataGroup:2024cfk}. This state 
was observed as a structure in the $B^{\pm}K^{\mp}$ invariant mass spectrum using a 
sample of proton-proton collision data collected with the LHCb detector at center-of-mass 
energies of $\sqrt{s}=7$, 8, and 13~TeV, corresponding to a total integrated luminosity 
of $9~\mathrm{fb}^{-1}$. The measured parameters, including the mass and width of the $B_{sJ}(6114)$ state, are given as
\begin{equation}
  \begin{aligned}
    m &= 6114 \pm 3 \, \text{(stat)} \pm 5 \, \text{(syst)} \, \text{MeV}, \\
    \Gamma &= 66 \pm 18 \, \text{(stat)} \pm 21 \, \text{(syst)} \, \text{MeV}.
  \end{aligned}
\end{equation}
However, its spin-parity quantum numbers have not yet been unambiguously determined~\cite{ParticleDataGroup:2024cfk}.

To clarify the spin-parity quantum numbers and internal structure of the $B_{sJ}(6114)$ state, various theoretical approaches have been proposed. Within the framework of the screened potential model, it was interpreted as a mixed state of the $2^3S_1$ and $1^3D_1$ configurations~\cite{Patel:2022hhl}. In contrast, the decay width analysis based on the non-relativistic quark potential model ruled out a significant $2S$ component and classified it as a member of the $1D(2^{-},3^{-})_{j_q=5/2}$ doublet~\cite{Chen:2022fye}. Furthermore, the $B_s(1^3D_1)$ assignment was suggested by the strong decay analysis within the effective chiral Lagrangian approach, which incorporates heavy-quark spin-flavor symmetry and light-quark chiral symmetry~\cite{Gandhi:2022nnk}. This interpretation was further supported by the Heavy Quark Effective Theory (HQET) analysis in Ref.~\cite{Garg:2023hvp}, where the calculated strong decay widths for different $J^P$ assignments were compared with the experimental measurement, excluding the $J^P=2^-$ and $J^P=3^-$ possibilities. More recently, the relativistic flux-tube model also favored a $1D$-wave excitation of the $B_s$ meson for the $B_{sJ}(6114)$ state~\cite{Jakhad:2025ejc}.

The above studies indicate that the $B_{sJ}(6114)$ state is likely to be a $1D$ $b\bar{s}$ excitation within the conventional quark-model framework. Nevertheless, its spin-parity assignment remains controversial, with possible candidates including $J^P=1^-$, $2^-$, and $3^-$~\cite{Chen:2022fye,Gandhi:2022nnk,Garg:2023hvp}. Such discrepancies may arise from the model dependence of different theoretical approaches, or alternatively suggest that the $B_{sJ}(6114)$ state may not be a conventional $1D$ $b\bar{s}$ excitation but instead possess a non-conventional internal structure. In this work, we favor the latter possibility, namely that the $B_{sJ}(6114)$ state may have a hadronic molecular structure. Indeed, within the hidden-gauge approach, the authors of Ref.~\cite{Sanchez-Illana:2026cyv} investigated the $B^{(*)}\bar{K}^{*}$ interaction and suggested that the $B_{sJ}(6114)$ state could be interpreted as a $S$-wave $B^{(*)}\bar{K}^{*}$ molecular state.  However, the mass spectrum alone from the $B^{(*)}\bar{K}^{*}$ interaction cannot unambiguously confirm the molecular nature of the $B_{sJ}(6114)$ state. Therefore, the decay properties should also be investigated within the molecular framework, as they provide crucial information for identifying its possible molecular structure.

In this work, we explore the possibility of interpreting the $B_{sJ}(6114)$ state as a $B^{(*)}\bar{K}^{*}$ molecular state  by systematically investigating both its mass spectrum and decay properties.  The binding energies $E$ of the $B^{(*)}\bar{K}^{*}$ systems are obtained by solving the nonrelativistic Schrödinger equation
\begin{equation}
\left[-\frac{1}{2\mu} \left( \nabla^2_r - \frac{L(L+1)}{r^2} \right) + V(r) \right] \psi(\vec{r}) = E \, \psi(\vec{r}),
\label{eq:schrodinger}
\end{equation}
within the Gaussian expansion method~\cite{Hiyama:2003cu}.  $\mu $ is the reduced mass of the system, and the radial Laplacian is given by $
\nabla_r^2 = \frac{1}{r^2} \frac{\partial}{\partial r} \left( r^2 \frac{\partial}{\partial r} \right)$.
Here, the interaction potentials $V(r)$ are derived from the effective Lagrangians within the one-boson-exchange (OBE) model. Without imposing any prior assumption on the orbital angular momentum $L$, we systematically investigate the possible spin-parity configurations of the $B^{(*)}\bar{K}^{*}$ molecular states, with the details summarized in Table~\ref{tab-1}.  The favored spin-parity assignment is determined by comparing the theoretical decay widths with the experimental data, where the decay widths are calculated within the framework developed in Ref.~\cite{Chen:2021tip} using the binding energies and wave functions obtained from solving Eq.~\eqref{eq:schrodinger} as inputs.
\begin{table}[htbp]
\centering
\caption{Possible quantum numbers $J^P$ for $B^{(*)}K^*$ systems involved in our calculation. We also take into account the coupling among different partial waves.}\label{tab-1}
\begin{tabular}{c c c c c}
\hline
$J^P$& $B^*K^*$ & $BK^*$ \\ \hline
$0^+$ & $| {^1S_0} \sim  {^5D_0} \rangle$ &  \\
$0^-$ & $| {^3P_0}  \rangle$ & $| {^3P_0}  \rangle$ \\
$1^+$ & $| {^3S_1} \sim  {^3D_1}\sim  {^5D_1} \rangle$ & $|{^3S_1} \sim  {^3D_1}\rangle$ \\
$1^-$ & $| {^1P_1} \sim  {^3P_1}\sim  {^5P_1}\sim  {^5F_1} \rangle$ & $|{^3P_1}\rangle$ \\
$2^+$ & $| {^5S_2} \sim  {^1D_2}\sim  {^3D_2} \sim  {^5D_2}\rangle$ & $|{^3D_2}\rangle$ \\
$2^-$ & $| {^3P_2} \sim  {^5P_2}\sim{^3F_2} \sim  {^5F_2} \rangle$ & $| {^3P_2}\sim ^3F_2\rangle$ \\
$3^+$&$| {^3D_3} \sim  {^5D_3} \rangle$&$|  {^3D_3} \rangle$\\
$3^-$&$| {^1F_3} \sim  {^3F_3}\sim{^5P_3} \sim  {^5F_3} \rangle$&$|  {^3F_3} \rangle$\\
\hline
\end{tabular}
\end{table}

This paper is organized as follows. In Sec.~\ref{sec:level2}, we present the theoretical formalism. The numerical results and related discussions are given in Sec.~\ref{sec:level3}, followed by a summary and conclusions in the last section.

\section{\label{sec:level2}Formalism and ingredients}
\begin{figure}[http]
\begin{center}
\includegraphics[scale=0.78]{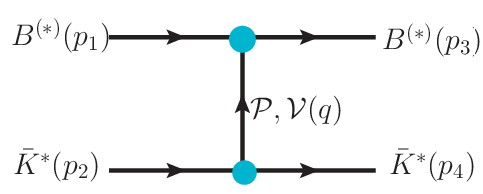}
\caption{Feynman diagrams for the processes $B^{(*)}\bar{K}^*$ $\to$ $B^{(*)}\bar{K}^*$ via the exchange of vector mesons $\rho,\omega$ and pseudoscalar mesons $\pi,\eta$. The corresponding momenta, $p_1,p_2,p_3,p_4$, are also indicated.}
\label{fig:combined}
\end{center}
\end{figure}

To obtain the bound-state energies and wave functions by solving the Schr\"odinger equation, we need first construct the interaction potential $V(r)$. In the present work, the potential is derived within the one-boson-exchange (OBE) framework based on the Feynman diagrams shown in Fig.~\ref{fig:combined}. The exchanged bosons include the pseudoscalar mesons ${\cal{P}}=\pi,\eta$ and the vector mesons ${\cal{V}}=\rho,\omega$.
The interaction potentials are obtained from the scattering amplitudes associated with these exchange processes. To this end, we employ the effective Lagrangians for the relevant meson-exchange vertices, which respect heavy-quark symmetry, chiral symmetry, and SU(2)-flavor symmetry~\cite{wise1992,yan1992,grinstein1992,casalbuoni1997,li2012,li2013}.    
\begin{align}
\mathcal{L}_{B^{(*)}B^{(*)}P}=&\frac{2g}{f_\pi}(BB^{*\dagger}_\lambda+B^*_\lambda B^\dagger)\partial^\lambda P\nonumber\\
&+i\frac{2g}{f_\pi}\nu^\mu\epsilon_{\mu\nu\alpha\beta}B^{*\nu}B^{*\alpha\dagger}\partial^\beta P\label{eq3},\\
\mathcal{L}_{B^{(*)}B^{(*)}V}&= \sqrt{2}\beta g_VBB^\dagger\nu\cdot V-2\sqrt{2}\lambda g_V\epsilon_{\mu\nu\alpha\beta}\nonumber\\
&\times\nu^\mu(BB^{*\nu\dagger}+B^{*\nu}B^{\dagger})\partial^\alpha V^\beta\nonumber\\
    &-\sqrt{2}\beta g_VB^{*}B^{*\dagger}\nu\cdot V\nonumber\\
    &-i2\sqrt{2}\lambda g_VB^{*\mu}B^{*\nu\dagger}(\partial_\mu V_\nu-\partial_\nu V_\mu),\\
    \mathcal{L}_{\bar{K}^{(*)}\bar{K}^{(*)}P}=&\frac{2g'}{f'_\pi}(KK^{*\dagger}_\lambda+K^*_\lambda K^\dagger)\partial^\lambda P\nonumber\\
    &+i\frac{2g'}{f_\pi}\nu'^{\mu}\epsilon_{\mu\nu\alpha\beta}\bar{K}^{*\nu}\bar{K}^{*\alpha\dagger}\partial^\beta P,\\
    \mathcal{L}_{\bar{K}^{(*)}\bar{K}^{(*)}V}=& \sqrt{2}\beta 'g_V'\bar{K}\bar{K}^\dagger\nu'\cdot V-2\sqrt{2}\lambda' g_V'\epsilon_{\mu\nu\alpha\beta}\nonumber\\    &\times\nu'^\mu(\bar{K}\bar{K}^{*\nu\dagger}+\bar{K}^{*\nu}\bar{K}^{\dagger})\partial^\alpha V^\beta\nonumber\\
    &-\sqrt{2}\beta' g'_V\bar{K}^{*}\bar{K}^{*\dagger}\nu'\cdot V\nonumber\\
    &-i2\sqrt{2}\lambda' g'_V\bar{K}^{*\mu}\bar{K}^{*\nu\dagger}(\partial_\mu V_\nu-\partial_\nu V_\mu)\label{eq6},
\end{align}
where $g=0.59$, $g'=1.12$, and $f_\pi=132~\mathrm{MeV}$ is the pion decay constant. 
The parameter $\beta'=0.835$ is determined within the hidden-gauge symmetry framework for vector mesons~\cite{molina2010}. 
The couplings $g_V=g'_V=m_\rho/f_\pi=5.8$ and $\beta=0.9$ are fixed by the vector meson dominance model, while 
$\lambda=\lambda'=0.56~\mathrm{GeV}^{-1}$ is obtained by comparing the form factors calculated from light-cone sum rules and lattice QCD results~\cite{isola2003,bando1988}. 
Here, $\epsilon_{\mu\nu\alpha\beta}$ denotes the Levi-Civita tensor with $\epsilon_{0123}=1$. 
The field matrices ${\cal{P}}$ and ${\cal{V}}$ in the $SU(2)$ flavor space are given as
\begin{align}
  {\cal{P}}&=\begin{pmatrix} 
\frac{\pi^0}{\sqrt{2}}+\frac{\eta}{\sqrt{6}}&\pi^+\\
\pi^-&-\frac{\pi^0}{\sqrt{2}}+\frac{\eta}{\sqrt{6}}
  \end{pmatrix},
\end{align}
\begin{align}
 {\cal{V}}&=\begin{pmatrix} 
\frac{\rho^0}{\sqrt{2}}+\frac{\omega}{\sqrt{2}}&\rho^+\\
\rho^-&-\frac{\rho^0}{\sqrt{2}}+\frac{\omega}{\sqrt{2}}
  \end{pmatrix}.
\end{align}

Since hadrons are extended objects rather than pointlike particles, phenomenological form factors are introduced to account for their finite-size effects. For the exchanged seudoscalar mesons ${\cal{P}}$ and vector mesons ${\cal{V}}$, we adopt the commonly used monopole form factor~\cite{Xie:2015zga,He:2011jp},
\begin{align}
{\cal{F}}(q^2)=\frac{\Lambda^2-m^2}{\Lambda^2-q^2},
\end{align}
where $q_i$ and $m_i$ denote the four-momentum and mass of the exchanged mesons, respectively. The cutoff parameter $\Lambda_i$ characterizes the size of the hadronic and is usually parameterized in a form close to the mass of the exchanged meson,
\begin{equation}
\Lambda_i=m_i+\alpha\Lambda_{\rm QCD},
\end{equation}
where $\Lambda_{\rm QCD}=220~\mathrm{MeV}$ is the QCD energy scale. The dimensionless parameter $\alpha$ reflects the nonperturbative QCD effects at low energies and cannot be determined from first principles within the present phenomenological framework. Instead, it is usually constrained by phenomenological analyses and experimental observables. In previous studies, a relatively broad range of values for $\alpha$ has been adopted depending on the specific hadronic systems and processes under investigation. For example, values of $\alpha=0.5$--$5.0$ were found to provide a reasonable description of the decay widths of the hidden-charm molecular candidates $X(3872)\to J/\psi\,\rho$ and $Y(3940)\to J/\psi\,\omega$~\cite{Liu:2006df}, and similar values have also been used in other molecular studies~\cite{Colangelo:2002mj,Meng:2007cx}. By fitting the $e^+e^-\to\bar{p}p\pi^0$ data at $\sqrt{s}=3.773~\mathrm{GeV}$, a constraint of $\alpha=6.2\pm3.5$ was obtained~\cite{Xu:2015qqa}. In addition, the experimental branching ratios of $\psi(4040)\to J/\psi\,\eta$ and $\psi(4160)\to J/\psi\,\eta$ can be reproduced with a smaller range of $\alpha=0.53$--$1.20$~\cite{Chen:2012nva}.  In this work, we therefore investigate the possible bound states by scanning $\alpha$ over the range $0.5$--$9.7$, which covers the values commonly adopted in previous studies.

By combining all the pieces together, the scattering amplitudes corresponding to the diagrams in Fig.~\ref{fig:combined} can be obtained as follows:
\begin{align}
\mathcal{M}_{\rho^0/\omega}(&B\bar{K}^*\to B\bar{K}^*)=-2\beta\beta'g_Vg'_V(\epsilon_2\cdot\epsilon_4^\dagger)\mathcal{F}^2_{\rho^0/\omega}\\  
    &\times{}C_{\rho^0/\omega}\frac{i(-g^{\mu\nu}+q^\mu q^\nu/m^2)}{q^2-m^2},\\
\mathcal{M}_{\rho^0/\omega}(&B^*\bar{K}^*\to B^*\bar{K}^*)=[2\beta\beta'g_Vg'_V(\epsilon_1\cdot\epsilon_3^\dagger)(\epsilon_2\cdot\epsilon_4^\dagger)\nonumber\\
&+8\lambda\lambda'g_Vg_V^{'}\epsilon_1^\alpha\epsilon_3^{\dagger\beta}(q_\beta g_{\mu\alpha}-q_\alpha g_{\mu\beta})\epsilon_2^\theta\epsilon_4^{\dagger\lambda}\nonumber\\
&\times(q_\theta g_{\nu\lambda}-q_\lambda g_{\nu\theta})]\mathcal{F}^2_{\rho^0/\omega}\nonumber\\
&\times{}C_{\rho^0/\omega}\frac{i(-g^{\mu\nu}+q^\mu q^\nu/m^2)}{q^2-m^2},\\
\mathcal{M}_{\pi/\eta}(&B^*\bar{K}^*\to B^*\bar{K}^*)=-\frac{4gg'}{f_\pi^2}\epsilon_{\mu\nu\alpha\theta}\nu^{\theta}\epsilon_1^\mu\epsilon_3^{\dagger\nu}q^\alpha
\end{align}
\begin{align}
&\times\epsilon_{\beta\lambda\tau\eta}\nu'^{\eta}\epsilon_2^\beta\epsilon_4^{\dagger\lambda}q^\tau\mathcal{F}^2_{\pi/\eta}C_{\pi/\eta}\frac{i}{q^2-m^2},\\
\mathcal{M}_{\rho^0/\omega}(&B\bar{K}^{*} \to B^{*}\bar{K}^{*})=8i\lambda\lambda'g_vg'_v\epsilon_{\mu\nu\alpha\beta}\times\nu^\mu\epsilon_3^{\nu\dagger}q^\alpha\nonumber\\
&\times{}(\epsilon_2^\theta\epsilon_{4\tau}^{\dagger}q_\theta i(-g^{\beta\tau}+q^\beta q^\tau/m^2)-\epsilon_{2\theta}\epsilon_4^{\dagger\tau} q_\tau i\nonumber\\
&\times{}(-g^{\beta\theta}+q^\beta q^\theta/m^2))\mathcal{F}^2_{\rho^0/\omega}C_{\rho^0/\omega}\frac{1}{q^2-m^2},\\
\mathcal{M}_{\pi/\eta}(&B\bar{K}^{*} \to B^{*}\bar{K}^{*})= \frac{4 i g g'}{f_{\pi}^{2}} \, \epsilon_{3\lambda}^{\dagger} \, q^{\lambda}  \, \nonumber\\
&\times\epsilon_{\mu\nu\alpha\beta} \, \nu'^\mu\epsilon_{2}^{\nu} \, \epsilon_{4}^{\alpha\,\dagger} \, q^{\beta} \, \mathcal{F}_{\pi/\eta}^{2} \, C_{\pi/\eta}\frac{i}{q^{2} - m^{2}},
\end{align}
where $m$ and $q$ are the mass and momentum of the exchanged particles. The isospin factors are
$C_{\pi}=-\frac{3}{2}$,
$C_{\rho}=-\frac{3}{2}$,
$C_{\eta}=\frac{1}{6}$,
and
$C_{\omega}=\frac{1}{2}$.
The polarization vector $\epsilon_{\mu}(s,p)$ of the spin-1 vector meson $K^{*}$ or $B^{*}$ can be chosen as
\begin{align}
\epsilon_\mu(s=0)=\begin{pmatrix}0\\0\\0\\1\end{pmatrix},
\epsilon_\mu(s=\pm)=\frac{1}{\sqrt{2}}\begin{pmatrix}0\\\mp1\\-i\\0\end{pmatrix}.
\end{align}

With the scattering amplitudes obtained from the effective Lagrangians, the interaction potentials for the $B^{(*)}K^{*}\to B^{(*)}K^{*}$ processes are derived in momentum space within the Breit approximation, which are given by
\begin{equation}
\mathcal{V}(\vec{q})=-\frac{\mathcal{M}}{\sqrt{\prod_i 2m_i\prod_f 2m_f}} .
\end{equation}
Here, $\mathcal{M}$ denotes the corresponding scattering amplitude, while $m_i$ and $m_f$ represent the masses of the initial and final particles, respectively. The coordinate-space potential, which can be directly employed in the Schrödinger equation~\eqref{eq:schrodinger} for numerical calculations, is obtained by performing the Fourier transformation of the momentum-space potential. For completeness, the Fourier transform relations required in this work are summarized as follows:
\begin{align}
    &\mathcal{F}r\left\{\frac{1}{\vec{q}^{\,2}+m^2}
    \left(\frac{\Lambda^2-m^2}{\Lambda^2+\vec{q}^{\,2}}\right)^2\right\}
    =Y_1(\Lambda,m,r), \\
    &\mathcal{F}r\left\{\frac{\vec{q}^{\,2}}{\vec{q}^{\,2}+m^2}
    \left(\frac{\Lambda^2-m^2}{\Lambda^2+\vec{q}^{\,2}}\right)^2\right\}
    =-\nabla_r^2Y_1(\Lambda,m,r),\\
     &\mathcal{F}r\left\{\frac{(\vec{A}\cdot\vec{q})(\vec{B}\cdot\vec{q})}
    {\vec{q}^{\,2}+m^2}
    \left(\frac{\Lambda^2-m^2}{\Lambda^2+\vec{q}^{\,2}}\right)^2\right\}=-\frac{1}{3}(\vec{A}\cdot\vec{B})\nabla_r^2\nonumber\\
    &\times{}Y_1(\Lambda,m,r)-\frac{1}{3}S(r,\vec{A},\vec{B})
    \left(r\frac{\partial}{\partial r}\frac{1}{r}
    \frac{\partial}{\partial r}Y_1(\Lambda,m,r)\right),
\end{align}
where \(Y_1(\Lambda,m,r)\) is the regulated Yukawa-type function,
\begin{align}
Y_1(\Lambda,m,r)
=&\frac{1}{4\pi r}\left(e^{-mr}-e^{-\Lambda r}\right)
-\frac{\Lambda^2-m^2}{8\pi\Lambda}e^{-\Lambda r},
\nonumber
\end{align}
and
\[
S(r,\vec{A},\vec{B})=
3\frac{(\vec{r}\cdot\vec{A})(\vec{r}\cdot\vec{B})}{r^2}
-\vec{A}\cdot\vec{B}
\]
represents the tensor operator.

The remaining ingredients required for the numerical calculations are the tensor operators and the matrix elements of other relevant operators. We collectively denote these operators by ${\cal{Z}}$, and their explicit forms are summarized in Tables~\ref{tab-2},~\ref{tab-3}, and~\ref{tab-4}. 
To evaluate their matrix elements, we introduce the spin-orbit wave functions characterized by the spin angular momentum $S$, orbital angular momentum $L$, and total angular momentum $J$ of the systems under consideration. The general form of these wave functions can be expressed as
\begin{align}
&|B\bar{K}^*(^{2S+1}L_J)\rangle
=\sum_{m_S,m_L} C^{J,M}_{S,m_S;L,m_L}\,
\chi_{m_S}|Y_{L,m_L}\rangle,\\
&|B^*\bar{K}^*(^{2S+1}L_J)\rangle
=\sum_{m_S,m_L} C^{J,M}_{S,m_S;L,m_L}\,
\chi_{m_S}|Y_{L,m_L}\rangle,
\end{align}
where $C^{J,M}_{S,m_S;L,m_L}$ denotes the Clebsch--Gordan coefficient, and $|Y_{L,m_L}\rangle$ represents the spherical harmonic function. The quantities $M$, $m_S$, and $m_L$ are the projections of the total angular momentum $J$, spin angular momentum $S$, and orbital angular momentum $L$, respectively, along the $z$ axis. With these spin-orbit wave functions, the matrix elements of the operators, $\langle {}^{2S'+1}L'_{J'}|{\cal{Z}}|{}^{2S+1}L_J\rangle$,
can be evaluated explicitly, and the corresponding results are presented in Tables~\ref{tab-2},~\ref{tab-3}, and~\ref{tab-4}.

\begin{sidewaystable}
\captionsetup{font=scriptsize, skip=2pt}
\vfill
\scriptsize
\caption{The matrix elements of two-body interaction operators for $B^{(*)}\bar{K}^*$ systems.}\label{tab-2}
\begin{tabular}{ccccc}
\toprule
\noindent
{$B^{(*)}K^{*}$\textrightarrow$B^{(*)}K^{*}$} & $0^+$ & $0^-$ & $1^+$ & $1^-$\\
$z$ & $(^{1}S_0, ^{5}D_0)$ & $(^{3}P_0)$ & 
$(^{3}S_1, ^{3}D_1, ^{5}D_1)$ & $(^{1}P_1, ^{3}P_1, ^{5}P_1,^{5}F_1)$ \\
\midrule
$(\vec{\epsilon}_1\vec{\epsilon}_2)(\vec{\epsilon}_3\vec{\epsilon}_4)$&$\begin{pmatrix} 1 & 0 \\ 0 & 1/5 \end{pmatrix}$&1/3&$\begin{pmatrix} 1 & 0 &0\\ 0 & 2/5&0\\0&0&0 \end{pmatrix}$&$\begin{pmatrix} 1 & 0&0&0 \\ 0 &0&0&0\\0&0&2/5&0\\0&0&0&9/35 \end{pmatrix}$\\

$(\vec{\epsilon}_2\vec{\epsilon}_3)(\vec{\epsilon}_1\vec{\epsilon}_4)$&$\begin{pmatrix} 1/3 & 0 \\ 0 & 8/15 \end{pmatrix}$&0&$\begin{pmatrix} 0 & 0 &0\\ 0 & 0&0\\0&0&4/5 \end{pmatrix}$&$\begin{pmatrix} 1/3 & 0&2/3\sqrt{5}&0 \\ 0 &0&0&0\\2/3\sqrt{5}&0&4/15&0\\0&0&0&16/35 \end{pmatrix}$\\

$(\vec{\epsilon}_2\vec{\epsilon}_4)(\vec{\epsilon}_1\vec{\epsilon}_3)$&$\begin{pmatrix} 1 & 0 \\ 0 & 1\end{pmatrix}$&1&$\begin{pmatrix} 1 & 0 &0\\ 0 & 1&0\\0&0&1 \end{pmatrix}$&$\begin{pmatrix}1 & 0&0&0 \\ 0 &1&0&0\\0&0&1&0\\0&0&0&1 \end{pmatrix}$\\

$(\vec{\epsilon}_1\times\vec{\epsilon}_3)(\vec{\epsilon}_2\times\vec{\epsilon}_4)$&$\begin{pmatrix} 2/3 & 0 \\ 0 & -1/3\end{pmatrix}$&1/3&$\begin{pmatrix} 1 & 0 &0\\ 0 & 2/3&0\\0&0&-4/5 \end{pmatrix}$&$\begin{pmatrix}2/3 & 0&-2/3\sqrt{5}&0 \\ 0 &0&0&0\\-2/3\sqrt{5}&0&2/15&0\\0&0&0&-1/5 \end{pmatrix}$\\

$S(r,\vec{\epsilon}_1\times\vec{\epsilon}_3,\vec{\epsilon}_2\times\vec{\epsilon}_4)$&$\begin{pmatrix} 0 & 2\sqrt{2}/15 \\ 2\sqrt{2}/15 & 4/15\end{pmatrix}$&4/15&$\begin{pmatrix} 0 & -2\sqrt{2}/5 &0\\ -2\sqrt{2}/5 & 8/35&0\\0&0&16/35 \end{pmatrix}$&$\begin{pmatrix}8/15 & 0&-8/15\sqrt{5}&6\sqrt{6/5}/35 \\ 0 &0&0&0\\-8/15\sqrt{5}&0&8/75&-6\sqrt{6}/175\\6\sqrt{6/5}/35&0&-6\sqrt{6}/175&8/175 \end{pmatrix}$\\

$\vec{\epsilon}_1\vec{\epsilon}_2 S(r,\vec{\epsilon}_3,\vec{\epsilon}_4)$&$\begin{pmatrix} 0 & -\sqrt{2}/5 \\ -\sqrt{2}/5 & 2/35\end{pmatrix}$&-2/15&$\begin{pmatrix} 0 & \sqrt{2}/5 &0\\ \sqrt{2}/5 & -4/35&0\\0&0&0 \end{pmatrix}$&$\begin{pmatrix}0 & 0&4/5\sqrt{5}&-9\sqrt{6/5}/35 \\ 0 &0&0&0\\4/5\sqrt{5}&0&4/25&-9\sqrt{6}/175\\-9\sqrt{6/5}/35&0&-9\sqrt{6}/175&12/175 \end{pmatrix}$\\

$\vec{\epsilon}_1\vec{\epsilon}_4 S(r,\vec{\epsilon}_2,\vec{\epsilon}_3)$&$\begin{pmatrix} 0 & -\sqrt{2}/15 \\ -\sqrt{2}/15 & 4/21\end{pmatrix}$&0&$\begin{pmatrix} 0 & 0 &0\\ 0 & 0&0\\0&0&8/35\end{pmatrix}$&$\begin{pmatrix}4/15 & 0&8/15\sqrt{5}&-6\sqrt{6/5}/35 \\ 0 &0&0&0\\8/15\sqrt{5}&0&16/75&-12\sqrt{6}/175\\-6\sqrt{6/5}/35&0&-12\sqrt{6}/175&16/175 \end{pmatrix}$\\
\midrule
\end{tabular}

\vfill
\end{sidewaystable}

\begin{sidewaystable}
\captionsetup{font=scriptsize, skip=2pt}
\vfill
\scriptsize
\caption{\scriptsize Continued}\label{tab-3}
\begin{tabular}{ccccc}
\toprule
\noindent
{$B^{(*)}K^{*}$\textrightarrow$B^{(*)}K^{*}$} & $2^+$ & $2^-$ & $3^+$ & $3^-$\\
$z$ & $(^1D_2,^3D_2,^5S_2,^5D_2)$ & $(^3P_2,^3F_2,^5P_2,^5F_2)$ & 
$(^3D_3,^5D_3)$ & $(^1F_3,^3F_3,^5P_3,^5F_3)$ \\
\midrule
$(\vec{\epsilon}_1\vec{\epsilon}_2)(\vec{\epsilon}_3\vec{\epsilon}_4)$&$\begin{pmatrix} 1 & 0&0&0 \\0& 0 & 0&0 \\0&0&1&0\\0&0&0&2/7\end{pmatrix}$&$\begin{pmatrix} 2/3 & 0&0&0 \\0& 3/7 & 0&0 \\0&0&0&0\\0&0&0&0\end{pmatrix}$&$\begin{pmatrix} 3/5 & 0 \\ 0 & 0\end{pmatrix}$&$\begin{pmatrix} 1 & 0&0&0 \\ 0 &0&0&0\\0&0&3/5&0\\0&0&0&4/15\end{pmatrix}$\\

$(\vec{\epsilon}_2\vec{\epsilon}_3)(\vec{\epsilon}_1\vec{\epsilon}_4)$&$\begin{pmatrix} 1/3 & 0&0&2/3\sqrt{7} \\0& 0 & 0&0 \\0&0&2/3&0\\2/3\sqrt{7}&0&0&16/21\end{pmatrix}$&$\begin{pmatrix} 0 & 0&0&0 \\0& 0 & 0&0 \\0&0&0&0\\0&0&0&5/7\end{pmatrix}$&$\begin{pmatrix} 0 & 0 \\ 0 & 1/5\end{pmatrix}$&$\begin{pmatrix} 1/3 & 0&0&2\sqrt{2/15}/3 \\ 0 &0&0&0\\0&0&2/5&0\\2\sqrt{2/15}/3&0&0&38/45\end{pmatrix}$\\

$(\vec{\epsilon}_2\vec{\epsilon}_4)(\vec{\epsilon}_1\vec{\epsilon}_3)$&$\begin{pmatrix} 1 & 0&0&0 \\ 0 & 1&0&0\\0&0&1&0\\0&0&0&1\end{pmatrix}$&$\begin{pmatrix} 1 & 0&0&0 \\ 0 & 1&0&0\\0&0&1&0\\0&0&0&1\end{pmatrix}$&$\begin{pmatrix} 1 & 0 \\ 0 & 1 \end{pmatrix}$&$\begin{pmatrix} 1 & 0&0&0 \\ 0 & 1&0&0\\0&0&1&0\\0&0&0&1\end{pmatrix}$\\

$(\vec{\epsilon}_1\times\vec{\epsilon}_3)(\vec{\epsilon}_2\times\vec{\epsilon}_4)$&$\begin{pmatrix} 2/3 & 0&0&-2/3\sqrt{7} \\ 0 &0&0&0\\0&0&1/3&0\\-2/3\sqrt{7}&0&0&-10/21\end{pmatrix}$&$\begin{pmatrix} 2/3 & 0&0&0 \\ 0 &3/7&0&0\\0&0&0&0\\0&0&0&-5/7\end{pmatrix}$&$\begin{pmatrix} 3/5 & 0\\ 0 & -1/5 \end{pmatrix}$&$\begin{pmatrix}2/3 & 0&0&-2\sqrt{2/15}/3 \\ 0 &0&0&0\\0&0&1/5&0\\-2\sqrt{2/15}/3&0&0&-26/45 \end{pmatrix}$\\

$S(r,\vec{\epsilon}_1\times\vec{\epsilon}_3,\vec{\epsilon}_2\times\vec{\epsilon}_4)$&$\begin{pmatrix} 8/21 & 0&2\sqrt{2/5}/3&-8/21\sqrt{7} \\ 0 & 0&0&0\\2\sqrt{2/5}/3&0&0&-2\sqrt{2/35}/3\\-8/21\sqrt{7}&0&-2\sqrt{2/35}/3&8/21\end{pmatrix}$&$\begin{pmatrix} 8/15 & -6\sqrt{6}/35 &0&0\\ -6\sqrt{6}/35 & 8/35&0&0\\0&0&0&0 \end{pmatrix}$&$\begin{pmatrix}12/35&0\\0&4/35\end{pmatrix}$&$\begin{pmatrix}16/45 & 0&6\sqrt{2/35}/5&-16\sqrt{2/15}/45 \\ 0 &0&0&0\\6\sqrt{2/35}/5&0&4/25&-4\sqrt{3/7}/25\\-16\sqrt{2/15}/45&0&-4\sqrt{3/7}/25&32/675 \end{pmatrix}$\\

$\vec{\epsilon}_1\vec{\epsilon}_2 S(r,\vec{\epsilon}_3,\vec{\epsilon}_4)$&$\begin{pmatrix} 0&0 & -\sqrt{2/5}&4/7\sqrt{7} \\ 0 & 0&0&0\\-\sqrt{2/5}&0&0&-\sqrt{2/35}\\4/7\sqrt{7}&0&\sqrt{2/35}&4/49\end{pmatrix}$&$\begin{pmatrix} -4/15 & 3\sqrt{6}/35 &0&0\\ 3\sqrt{6}/35 & -4/35&0&0\\0&0&0 &0\\0&0&0&0\end{pmatrix}$&$\begin{pmatrix}-6/35&0\\0&0 \end{pmatrix}$&$\begin{pmatrix}0 & 0&-9\sqrt{2/35}/5&8\sqrt{2/15}/15\\0&0&0&0\\-9\sqrt{2/35}/5&0&6/25&-6\sqrt{3/7}/25\\8\sqrt{2/15}/15&0&-6\sqrt{3/7}/25&16/225 \end{pmatrix}$\\

$\vec{\epsilon}_1\vec{\epsilon}_4 S(r,\vec{\epsilon}_2,\vec{\epsilon}_3)$&$\begin{pmatrix} 4/21&0 & -2\sqrt{2/5}/3&8/21\sqrt{7}\\ 0 & 0&0&0\\-2\sqrt{2/5}/3&0&0&-4\sqrt{2/35}/3\\8/21\sqrt{7}&0&-4\sqrt{2/35}/3&40/147\end{pmatrix}$&$0$&$\begin{pmatrix}0&0\\0&2/35\end{pmatrix}$&$\begin{pmatrix}8/45 & 0&-6\sqrt{2/35}/5&16\sqrt{2/15}/45\\0&0&0&0\\-6\sqrt{2/35}/5&0&8/25&-8\sqrt{3/7}/25\\16\sqrt{2/15}/45&0&-8\sqrt{3/7}/25&64/675 \end{pmatrix}$\\
\midrule
\end{tabular}
\vfill
\end{sidewaystable}
\begin{sidewaystable}
\captionsetup{font=scriptsize, skip=2pt}
\vfill
\scriptsize
\caption{\scriptsize Continued}\label{tab-4}
\begin{tabular}{ccccc}
\toprule
$B^{(*)}K^{*}\rightarrow B^{(*)}K^{*}$ & $0^+$ & $0^-$ & $1^+$ & $1^-$ \\
$z$ & $(^{1}S_0, ^{5}D_0)$ & $(^{3}P_0)$ & 
$(^{3}S_1, ^{3}D_1, ^{5}D_1)$ & $(^{1}P_1, ^{3}P_1, ^{5}P_1,^{5}F_1)$ \\
\bottomrule
$i\vec\epsilon_3(\vec\epsilon_2\times\vec\epsilon_4)$&&$-\sqrt{2}/3$&$\begin{pmatrix} 0 & 0&0 \\ 0 & -3/5\sqrt{2}&-\sqrt{3/2}/5 \end{pmatrix}$&$\begin{pmatrix} 0&-1/\sqrt{2}&\sqrt{3/10} & 0 \end{pmatrix}$ \\
$iS(r,\vec\epsilon_3,\vec\epsilon_2\times\vec\epsilon_4)$&&$\sqrt{2}/3$&$\begin{pmatrix} 0&0 & 0 \\ -3/5&0 & -\sqrt{6}/5 \end{pmatrix}$&$\begin{pmatrix} \sqrt{6}/5&\sqrt{2}/5&-2\sqrt{6/5}/5&-9\sqrt{5}/25\end{pmatrix}$\\
$iS(r,\vec\epsilon_2,\vec\epsilon_3\times\vec\epsilon_4)$&&$-7\sqrt{2}/15$&$\begin{pmatrix} 0 & 3/5&-\sqrt{3}/5 \\ 3/5 &-9\sqrt{2}/70&3\sqrt{3/2}/35\end{pmatrix}$&$\begin{pmatrix} -\sqrt{6}/5&-\sqrt{2}/10&3\sqrt{3/10}/5&3\sqrt{5}/25 \end{pmatrix}$\\
$\vec\epsilon_2\vec\epsilon_4$&&1&$\begin{pmatrix}
    1&0\\0&1
\end{pmatrix}$&1\\
\midrule
{$B^{(*)}K^{*}$\textrightarrow$B^{(*)}K^{*}$} & $2^+$ & $2^-$ & $3^+$ & $3^-$\\
$z$ & $(^1D_2,^3D_2,^5S_2,^5D_2)$ & $(^3P_2,^3F_2,^5P_2,^5F_2)$ & 
$(^3D_3,^5D_3)$ & $(^1F_3,^3F_3,^5P_3,^5F_3)$ \\
\midrule
$i\vec\epsilon_3(\vec\epsilon_2\times\vec\epsilon_4)$&$\begin{pmatrix} 0&-1/\sqrt{2}&0&1/\sqrt{14} \end{pmatrix}$&$\begin{pmatrix} -\sqrt{2}/6&0&1/\sqrt{6}&0\\0&-2\sqrt{2}/7&0&-2/7 \end{pmatrix}$&$\begin{pmatrix}0&0\end{pmatrix}$&$\begin{pmatrix}0&-1/\sqrt{2}&0&\sqrt{3/10}\end{pmatrix}$ \\
$iS(r,\vec\epsilon_3,\vec\epsilon_2\times\vec\epsilon_4)$&$\begin{pmatrix} \sqrt{2}/7&-\sqrt{2}/7&1/\sqrt{5}&-6\sqrt{2/7}/7 \end{pmatrix}$&$\begin{pmatrix}-2\sqrt{2}/15&-\sqrt{3}/5&-\sqrt{2/3}/5&-\sqrt{6}/5\\-12\sqrt{3}/35&-2\sqrt{2}/35&12/35&-2/5 \end{pmatrix}$&$\begin{pmatrix}0&0\end{pmatrix}$&$\begin{pmatrix}2/15&\sqrt{2}/5&12\sqrt{2/35}/5&-14/15&\sqrt{2/15}\end{pmatrix}$\\
$iS(r,\vec\epsilon_2,\vec\epsilon_3\times\vec\epsilon_4)$&$\begin{pmatrix} -\sqrt{2}/7&\sqrt{2}/14&-1/\sqrt{5}&\sqrt{14}/98 \end{pmatrix}$&$\begin{pmatrix} 11\sqrt{2}/30&13\sqrt{3}/35&7\sqrt{6}/30&-3\sqrt{6}/35\\13\sqrt{3}/35&-2\sqrt{2}/35&3/35&4/35 \end{pmatrix}$&$\begin{pmatrix}0&0\end{pmatrix}$&$\begin{pmatrix}-2/15&\sqrt{2}/10&-9\sqrt{2/35}/5&\sqrt{30}/450\end{pmatrix}$\\
$\vec\epsilon_2\vec\epsilon_4$&1&$\begin{pmatrix}
    1&0\\0&1
\end{pmatrix}$&1&1\\
\midrule
\end{tabular}
\vfill
\end{sidewaystable}

Next, we introduce the formalism for calculating the decay widths. In this work, we focus on two decay channels of $B_{sJ}(6114)$, namely the experimentally observed $B_{sJ}(6114)\to \bar{K}B$ channel and the $B_{sJ}(6114)\to \bar{K}B^{*}$ channel, which has not yet been experimentally observed. The two-body decay width is related to the decay amplitude through
\begin{align}
\Gamma=
\frac{1}{2J+1}
\frac{|\mathbf{p}|}{8\pi m_{B_{sJ}}^2}
\left|\mathcal{M}(B_{sJ}\to \bar{K}B^{(*)})\right|^2,
\end{align}
where $|\mathbf{p}|$ denotes the three-momentum of the final-state particles in the rest frame of the initial state, $\mathcal{M}$ represents the corresponding decay amplitude, and $J$ is the total angular momentum of the $B_{sJ}(6114)$ state.

\begin{figure*}[http]
\begin{center}
\includegraphics[scale=0.65]{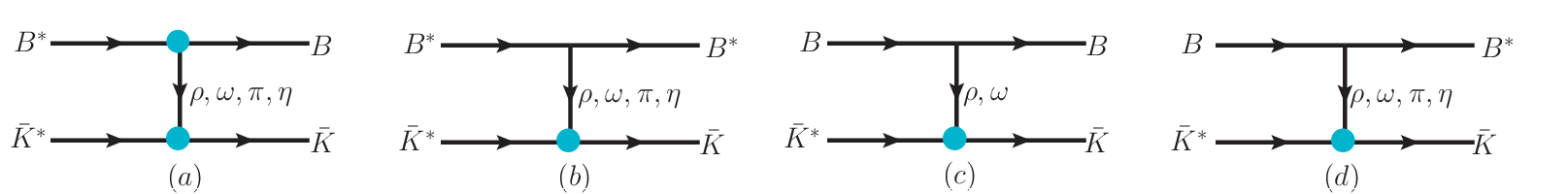}
\caption{Two-body strong decay diagrams for the $B_{sJ} (6114)$ as a coupled $B\bar{K}^*$/$B^*\bar{K}^*$ molecule at the hadron level.}
\label{fig:decay_processes}
\end{center}
\end{figure*}
Following Ref.~\cite{Chen:2021tip}, the decay amplitude of the $B_{sJ}(6114)$ state into the final state $\bar{K}+B^{(*)}$, within the $B^{(*)}\bar{K}^{*}$ molecular picture, can be expressed as the sum of the contributions from its $B\bar{K}^*$ and $B^*\bar{K}^*$ components:
\begin{align} &\langle \bar{K}B^{(*)}|B_{sJ}(6114)\rangle=\nonumber\\ &\langle \bar{K}B^{(*)}|\left(|B\bar{K}^*\rangle\langle B\bar{K}^*| + |B^*\bar{K}^*\rangle\langle B^*\bar{K}^*| \right) |B_{sJ}(6114)\rangle \nonumber\\ &= \int \frac{d^3k}{(2\pi)^3}d^3r\, e^{-i\mathbf{k}\cdot\mathbf{r}} \psi_{BK^*}(\mathbf{r}) \langle \bar{K}B^{(*)}|B\bar{K}^*\rangle \nonumber\\ &\quad+ \int \frac{d^3k}{(2\pi)^3}d^3r\, e^{-i\mathbf{k}\cdot\mathbf{r}} \psi_{B^*K^*}(\mathbf{r}) \langle \bar{K}B^{(*)}|B^*\bar{K}^*\rangle,\label{eq24} 
\end{align}
where $\psi_{BK^*}(\mathbf{r})$ and $\psi_{B^*K^*}(\mathbf{r})$ denote the coordinate-space wave functions of the $B\bar{K}^*$ and $B^*\bar{K}^*$ components, respectively. These wave functions are obtained by solving the coupled-channel Schrödinger equation described in the previous section. The corresponding matrix elements are given by
\begin{align}
\langle \bar{K}B^{(*)} | B_{sJ} \rangle &=
-\frac{\mathcal{M}\bigl(B_{sJ}\to K\bar{B}^{(*)}\bigr)}
{(2\pi)^{3/2}
\sqrt{2E_{B_{sJ}}}
\sqrt{2E_{\bar{K}}}
\sqrt{2E_{B^{(*)}}}},\\
\langle \bar{K}B^{(*)} | B^{(*)}\bar{K}^* \rangle
&=
-\frac{
\mathcal{M}\bigl(B^{(*)}\bar{K}^*\to \bar{K}B^{(*)}\bigr)}
{(2\pi)^{3/2}
\sqrt{2E_{B^{(*)}}}
\sqrt{2E_{\bar{K}^*}}
\sqrt{2E_{\bar{K}}}
\sqrt{2E_{\bar{B}^{(*)}}}},
\label{eq26}
\end{align}
where $E$ denotes the energy of the corresponding particle. Here,
$\mathcal{M}(B^{(*)}\bar{K}^*\to \bar{K}B^{(*)})$ denotes the transition amplitude for the $B^{(*)}\bar{K}^*\to \bar{K}B^{(*)}$ process. The relevant transition mechanisms involve the exchange of vector mesons $\rho$ and $\omega$ and pseudoscalar mesons $\pi$ and $\eta$, as illustrated in Fig.~\ref{fig:decay_processes}.
 Using the vertex interaction Lagrangians given in Eqs.~\eqref{eq3}-\eqref{eq6}, the decay amplitudes can be derived as
\begin{align}
   \mathcal{M}^V_{(a)B^*\bar{K}^*\to B\bar{K}} &=8\lambda g_v\lambda' g_v'\epsilon_{\lambda\mu\alpha\beta}\epsilon_{\tau\theta\gamma\sigma}\epsilon_1^\mu\epsilon_2^\theta q^{\alpha}q^{\gamma}\nonumber\\
   &\times\nu^\lambda\nu'^\tau\frac{-ig^{\sigma\beta}}{q^2-m^2_V}C_V,\\
\mathcal{M}^P_{(a)B^*\bar{K}^*\to B\bar{K}}&=\frac{4gg'}{f^2_\pi}\epsilon_1^\mu\epsilon_2^\nu q_{\mu}q_{\nu}\frac{i}{q^2-m_P^2}C_P,\\
\mathcal{M}^V_{(b)B^*\bar{K}^*\to B^*\bar{K}}&=8\lambda\lambda'g_vg'_v\epsilon_{\mu\nu\alpha\beta}\nu^\mu\epsilon_2^{\nu}q^\alpha(\epsilon_1^\theta\epsilon_{3\beta}^\dagger q_\theta \nonumber\\
&-\epsilon_{1\theta}\epsilon_3^{\tau\dagger} q_\tau g^{\beta\theta})\frac{1}{q^2-m_V^2}C_{V},\\
\mathcal{M}^P_{(b)B^*\bar{K}^*\to B^*\bar{K}}&=i\frac{4gg'}{f^2_\pi}\nu^{\mu}\epsilon_{\mu\nu\alpha\beta}\epsilon_1^\nu\epsilon_3^{\alpha\dagger}q^\beta\epsilon_2^\lambda q_\lambda\frac{i}{q^2-m_P^2}C_P,\\
  \mathcal{M}^V_{(c)B\bar{K}^*\to B\bar{K}}&=-4\beta g_v \lambda'g'_v\epsilon_{\lambda\mu\alpha\beta}\epsilon^\mu_2(iq)^\alpha\nu^\lambda\nu^\tau \frac{-ig^{\beta\tau}}{q^2-m^2_V}C_V,\\\nonumber
\mathcal{M}^V_{(d)B\bar{K}^*\to B^*\bar{K}}&=8\lambda g_v\lambda' g_v'\epsilon_{\lambda\mu\alpha\beta}\epsilon_{\tau\theta\gamma\sigma}\epsilon_3^{\mu\dagger}\epsilon_2^\theta(iq^\alpha)(-iq^\gamma)\\
    &\times\nu^\lambda\nu'^\tau\frac{-ig^{\sigma\beta}}{q^2-m^2_V}C_V,\\
\mathcal{M}^P_{(d)B\bar{K}^*\to B^*\bar{K}}&=\frac{4gg'}{f^2_\pi}\epsilon_2^\mu\epsilon_3^{\dagger\nu}q_\mu q_{\nu}\frac{i}{q^2-m_P^2}C_P.
\end{align}
$\mathcal{M}^{P(V)}_{i_1i_2\to f_1f_2}$corresponds to the amplitude of scattering of the process by exchanging the hadron $P(V)$.

\section{\label{sec:level3}results and discussions}
\begin{table*}[htbp]
\centering
\caption{Possible bound states of the $B^{(*)}\bar{K}^*$ system for different spin-parity assignments, with $\Lambda_i=m_i+220\alpha$ MeV. The coupled-channel results for $J^P=0^+$ are omitted because channel coupling has no effect in this sector. Here, $E$ denotes the calculated binding energy, while $P(\%)$ represents the percentage contribution of each partial wave to the molecular state. For coupled-channel systems, the channel composition is instead expressed directly in terms of $B\bar{K}^{(*)}(\%)$. All energies are given in MeV.}
\begin{tabular}{cccccccccc}
\toprule
$B\bar{K}^*$&State& $\alpha$ &E&P($\%$)&$B^*\bar{K}^*$&State&$\alpha$&E&P($\%$)\\
\midrule
&&&&&&&&&($^1S_0/^5D_0$)\\
&&&&&&$J^P=0^+$&0.5&-0.40&99.97/0.03\\
&&&&&&&0.6&-1.33&99.99/0.01\\
&&&&&&&0.7&-2.94&99.99/0.01\\
&$J^P=0^-$&1.6&-0.19&&&$J^P=0^-$&1.3&-0.51&\\
& &1.7&-0.84&&&&1.4&-1.80&\\
& &1.8&-1.68&&&&1.5&-3.49&\\
&&&&&&&&&($^1P_1/^3P_1/^5P_1/^5F_1$)\\
&$J^P=1^-$&1.6&-0.19&&&$J^P=1^-$&1.0&-0.18&83.32/0/16.66/0.02\\
& &1.7&-0.84&&&&1.1&-1.83&83.30/0/16.66/0.04\\
& &1.8&-1.68&&&&1.2&-4.23&83.25/0/16.65/0.10\\
&&&&($^3S_1/^3D_1$)&&&&&($^3S_1/^3D_1/^5D_1$)\\
&$J^P=1^+$&0.7&-0.30&100/0&&$J^P=1^+$&0.4&-0.16&99.75/0.25/0\\
& &0.8&-0.85&100/0&&&0.5&-0.91&99.88/0.12/0\\
& &0.9&-1.66&100/0&&&0.6&-2.44&99.95/0.05/0\\
&&&&($^3P_2/^3F_2$)&&&&&($^3P_2/^5P_2/^3F_2/^5F_2$)\\
&$J^P=2^-$&1.6&-0.19&100/0&&$J^P=2^-$&1.1&-0.73&99.64/0.36/0/0\\
& &1.7&-0.84&100/0&&&1.2&-2.52&99.83/0.17/0/0\\
& &1.8&-1.68&100/0&&&1.3&-4.96&99.90/0.10/0/0\\
&&&&&&&&&($^5S_2/^1D_2/^3D_2/^5D_2$)\\
&$J^P=2^+$&3.4&-0.53&&&$J^P=2^+$&0.7&-0.86&0.18/0/99.80/0.01\\
& &3.5&-1.31&&&&0.8&-2.14&0.20/0/99.79/0.01\\
& &3.6&-2.15&&&&0.9&-3.97&0.20/0/99.79/0.01\\
&&&&&&&&&($^1F_3/^3F_3/^5P_3/^5F_3$)\\
&$J^P=3^-$&8.3&-0.29&&&$J^P=3^-$&1.4&-0.35&0.60/0/99.34/0.06\\
&&8.4&-0.73&&&&1.5&-1.37&0.41/0/99.55/0.04\\
&&8.5&-1.18&&&&1.6&-2.69&0.30/0/99.67/0.03\\
&&&&&&&&&($^3D_3/^5D_3$)\\
&$J^P=3^+$&3.4&-0.53&&&$J^P=3^+$&2.0&-0.93&100/0\\
&&3.5&-1.31&&&&2.1&-3.33&100/0\\
&&3.6&-2.15&&&&2.2&-6.23&100/0\\
\bottomrule
$B\bar{K}^*\to B^*\bar{K}^*$&State& $\alpha$ &E&$B\bar{K}^*($\%$)$&$B^*\bar{K}^*($\%$)$\\
\midrule
&$J^P=0^-$&1.0&-0.20&44.87&55.13\\
&&1.1&-0.91&40.23&59.77\\
&&1.2&-2.03&37.11&62.89\\
&$J^P=1^-$&0.8&-0.18&49.87&50.13\\
&&0.9&-0.85&45.22&54.78\\
&&1.0&-2.03&40.34&59.66\\
&$J^P=1^+$&0.4&-0.17&0.01&99.99\\
&&0.5&-0.91&0.01&99.99\\
&&0.6&-2.46&0.01&99.99\\
&$J^P=2^-$&1.1&-1.81&16.88&83.12\\
&&1.2&-4.03&16.26&83.74\\
&&1.3&-6.96&15.56&84.44\\
&$J^P=2^+$&0.7&-0.94&$1.45$&98.55\\
&&0.8&-2.25&$1.26$&98.74\\
&&0.9&-4.13&$1.02$&98.98\\
&$J^P=3^-$&1.3&-0.01&0.01&99.99\\
&&1.4&-0.74&0.01&99.99\\
&&1.5&-1.82&0.01&99.99\\
&$J^P=3^+$&2.0&-1.38&1.36&98.64\\
&&2.1&-3.86&1.23&98.77\\
&&2.2&-6.84&0.99&99.01\\
\bottomrule
\end{tabular}
\end{table*}

In Table~\uppercase\expandafter{\romannumeral 6}, we present the properties of the bound states of the $B^{(*)}\bar{K}^*$ systems with $J^P=0^{\pm}, 1^{\pm}, 2^{\pm},$ and $3^{\pm}$.  For the $B\bar{K}^*$ system, the $B$ and $\bar{K}^*$ mesons have $J^P=0^-$ and $1^-$, respectively, and thus a molecular state with $J^P=0^+$ cannot be formed.  In contrast, the system can form a $J^P=0^-$ molecular state, corresponding to a $P$-wave molecular configuration, with only the $^3P_0$ partial wave contributing to this state.  As can be seen from the table, this $P$-wave $B\bar{K}^*$ molecular state emerges at $\alpha=1.6$, and its binding energy gradually increases with increasing $\alpha$, indicating that the state becomes more deeply bound.

\begin{figure}
\begin{center}
\includegraphics[scale=0.6]{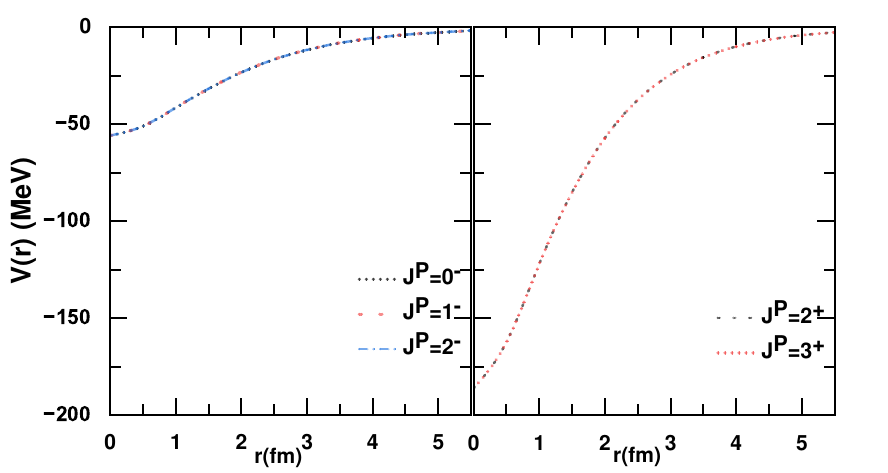}
\caption{The interaction potentials for the \(B\bar{K}^*\) system with \(J^P=0^-,\,1^-,\,2^\pm,\) and \(3^+\) at \(\alpha=1.6\).}
\label{fig:decay_processes-3}
\end{center}
\end{figure}

It is worth noting that another $P$-wave molecular state exists in the $J^P=1^-$ channel. Interestingly, its binding energy is identical to that of the $J^P=0^-$ state. Specifically, in both channels, the bound state first emerges at $\alpha=1.6$, with a binding energy of $E=0.19~\mathrm{MeV}$. When $\alpha$ is increased to $1.7$, the binding energy of both states increases to $E=1.68~\mathrm{MeV}$. This identical binding behavior can be attributed to the fact that the interaction potentials in these two channels are exactly the same, as shown in Fig.~\ref{fig:decay_processes-3}.

For the $B\bar{K}^{*}$ system in the $J^P=2^-$ channel, a similar bound-state behavior is also found, and the underlying mechanism is described as follows. This channel receives contributions from the ${}^3P_2$ and ${}^3F_2$ partial waves. Since the off-diagonal matrix element of the interaction potential between these two partial waves vanishes, they remain decoupled, and the resulting binding energies are identical to those obtained in the corresponding single-channel calculations. The ${}^3P_2$ component dominates the molecular state, accounting for nearly 100\% of its composition (see Table~\uppercase\expandafter{\romannumeral 6}), because the ${}^3F_2$ channel experiences a stronger central repulsive interaction than the ${}^3P_2$ channel. Moreover, the interaction potential in the ${}^3P_2$ channel is identical to those in the $J^P=0^-$ and $J^P=1^-$ channels, as shown in Fig.~\ref{fig:decay_processes-3}. We also find that the $\bar{K}^{*}B$ systems in the $J^P=2^+$ and $J^P=3^+$ channels, corresponding to two pure $D$-wave molecular states, exhibit identical binding properties because their interaction potentials are the sam (see Fig.~\ref{fig:decay_processes-3}).


Among all the channels considered, the $J^P=1^{+}$ and $J^P=3^{-}$ channels represent the most and least favorable cases, respectively, for the formation of bound states in the $\bar{K}^{*}B$ system. In the $J^P=1^{+}$ channel, the bound state emerges at the smallest value of $\alpha$, namely $\alpha=0.7$, with a binding energy of $E=0.30~\mathrm{MeV}$. Even when $\alpha$ is increased to $0.9$, the binding energy reaches only $E=1.66~\mathrm{MeV}$. In contrast, the bound state in the $J^P=3^{-}$ channel does not emerge until $\alpha$ reaches a much larger value of $8.3$, at which the binding energy is only $E=0.29~\mathrm{MeV}$. This behavior can be readily understood from the partial-wave contents of these two channels. The $J^P=1^{+}$ channel is dominated by the $S$-wave component, which is absent in all the other channels, whereas the $J^P=3^{-}$ channel contains only the $F$-wave component and experiences the strongest repulsive effect among all the channels.  Note that although the $\bar{K}B^{*}$ molecular state in the $J^P=2^{-}$ channel contains a ${}^3F_2$-wave component, which experiences the same centrifugal barrier as the $\bar{K}B^{*}$ molecular state in the $J^P=3^{-}$ channel, its dominant contribution comes from the ${}^3P_2$ wave. These results are consistent with the general expectation that $S$-wave molecular states are more readily formed than those dominated by higher partial waves. Moreover, molecular-state formation becomes increasingly difficult as the centrifugal barrier increases.

We now turn to the pure $\bar{K}^{*}B^{*}$ system. In the $J^P=0^+$ channel, which couples the ${}^1S_0$ and ${}^5D_0$ partial waves, a bound state emerges at $\alpha=0.5$ with a binding energy of $E=0.40~\mathrm{MeV}$. The binding energy increases to $1.33$ and $2.94~\mathrm{MeV}$ as $\alpha$ increases to $0.6$ and $0.7$, respectively. The $J^P=1^+$ molecular state is also dominated by the $S$ wave and has a similar critical value of $\alpha$. By contrast, the $J^P=2^+$ state is dominated by the ${}^3D_2$ partial wave, with the $S$-wave fraction increasing only from $0.18\%$ to $0.20\%$ as $\alpha$ varies from $0.7$ to $0.9$. The centrifugal barrier associated with the $D$ wave makes binding more difficult, and this state first appears at $\alpha=0.7$, with a binding energy of $0.86~\mathrm{MeV}$, requiring a larger $\alpha$ than the two $S$-wave-dominated states.

Despite its small fraction, the $S$-wave component may nevertheless play an important role in the formation of the $J^P=2^+$ state. To explore this possibility, we compare it with the pure $D$-wave state in the $J^P=3^+$ channel. The latter emerges only at $\alpha=2.0$, with a binding energy of $0.93~\mathrm{MeV}$, whereas the $J^P=2^+$ state appears at $\alpha=0.7$. This difference suggests that even a small $S$-wave component may facilitate the formation of a loosely bound state.

\begin{figure*}[http!]
\begin{center}
\includegraphics[scale=0.5]{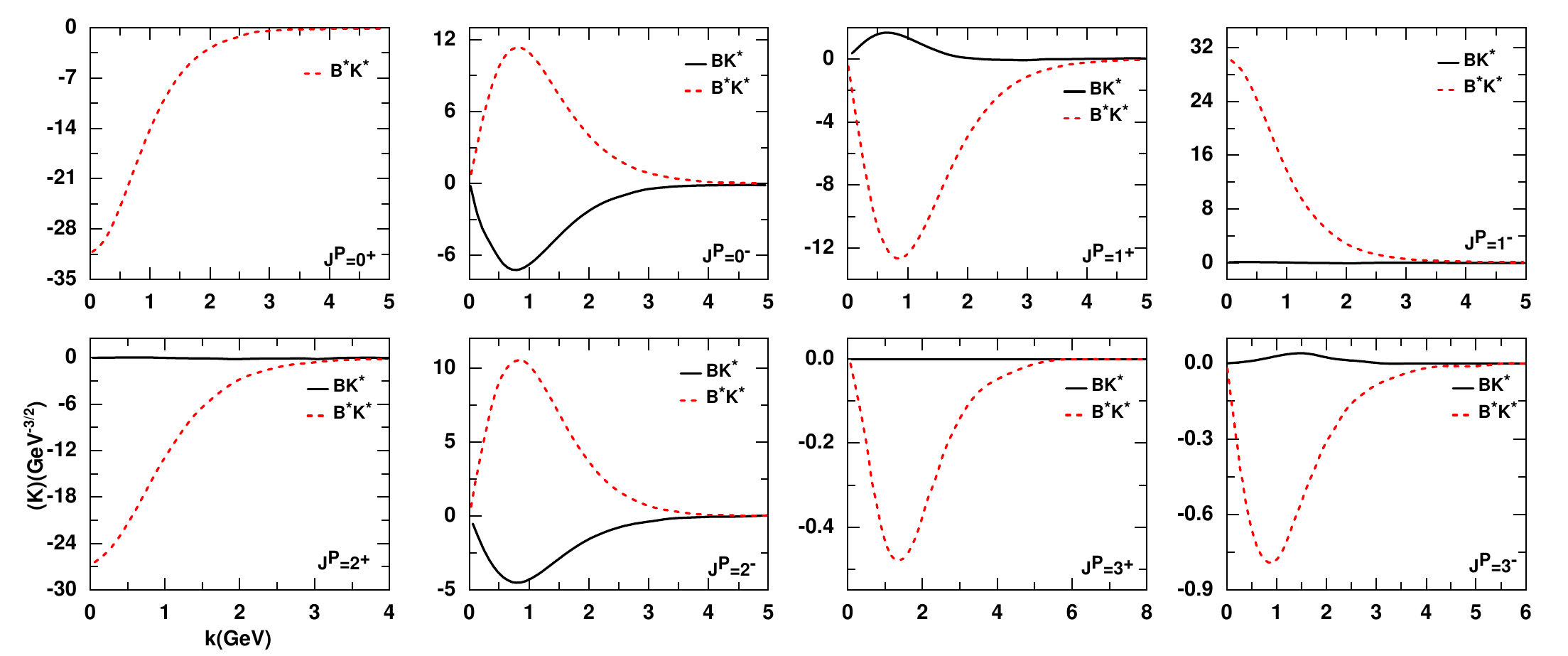}
\caption{The radial wave functions for the  $BK^*/B^*K^*$ molecule in the momentum space.  Here, its binding energy is E=-57 MeV.}
\label{fig:decay_processes}
\end{center}
\end{figure*}
A comparison with the $\bar{K}^{*}B$ system provides further insight. In its $J^P=0^-$ channel, the interaction involves only the ${}^3P_0$ partial wave. Although a $D$ wave generally encounters a stronger centrifugal barrier than a $P$ wave, the $D$-wave-dominated $J^P=2^+$ state in the $\bar{K}^{*}B^{*}$ system emerges at a smaller $\alpha$ than the pure $P$-wave $J^P=0^-$ state in the $\bar{K}^{*}B$ system. As shown in Table~\uppercase\expandafter{\romannumeral 6}, the latter appears at $\alpha=1.3$, with a binding energy of $0.51~\mathrm{MeV}$, compared with $\alpha=0.7$ and $E=0.86~\mathrm{MeV}$ for the former. This comparison further supports the possibility that even a small $S$-wave component can facilitate bound-state formation. One possible explanation is that the $S$-wave component interferes strongly with the other partial-wave components, thereby facilitating the formation of a bound state.

By comparing the other $P$-wave-dominated $\bar{K}^{*}B^{*}$ molecular states, we find that the more $F$-wave components are mixed in, the more difficult it is for the molecular state to form. The $J^P=1^-$, $2^-$, and $3^-$ channels involve four partial-wave components, namely $|{^1P_1} \sim {^3P_1} \sim {^5P_1} \sim {^5F_1}\rangle$, $|{^3P_2} \sim {^5P_2} \sim {^3F_2} \sim {^5F_2}\rangle$, and $|{^1F_3} \sim {^3F_3} \sim {^5P_3} \sim {^5F_3}\rangle$, respectively. These channels contain one, two, and three $F$-wave components, respectively. Their effects are directly reflected in the values of $\alpha$ at which bound states first emerge, as listed in Table~\uppercase\expandafter{\romannumeral 6}. As the number of mixed $F$-wave components increases, a larger value of $\alpha$ is required for a bound state to emerge, indicating that stronger meson-exchange attraction is needed to form the molecular state. Specifically, the $J^P=1^-$ state, which contains only one $F$-wave component, first appears as a bound state at $\alpha=1.0$, with a binding energy of $0.18$ MeV. In contrast, the $J^P=2^-$ and $3^-$ states, which contain two and three $F$-wave components, respectively, appear as bound states at $\alpha=1.1$  and $1.4$, with corresponding binding energies of $0.73$ MeV and $0.35$ MeV. 
Note that no molecular states are found in either the $J^P=2^-$ or $J^P=3^-$ channel at $\alpha=1.0$. The reason still is that, although these $F$-wave contributions are small, their interference with the $P$-wave components is strong.

After including the coupling between the $\bar{K}^{*}B$ and $\bar{K}^{*}B^{*}$ channels, the most notable change for molecular states with substantial contributions from both channels is that bound states that require relatively large cutoff parameters to emerge in single-channel calculations can form at considerably smaller values of $\alpha$ in the coupled-channel calculation. Taking the $J^P=0^{-}$ channel as an example, a bound state emerges at $\alpha=1.6$ in the single-channel $\bar{K}^{*}B$ calculation. In contrast, in the coupled $\bar{K}^{*}B/\bar{K}^{*}B^{*}$ calculation, a bound state already appears at $\alpha=1.0$, which is also smaller than the minimum value of $\alpha=1.3$ required for the $\bar{K}^{*}B^{*}$ system to form a bound state. The binding energy is approximately $0.20$ MeV, and the probabilities of the $\bar{K}^{*}B$ and $\bar{K}^{*}B^{*}$ components are approximately $45\%$ and $55\%$, respectively.  As $\alpha$ increases to 1.1 and 1.2, the binding energy increases to $0.91$ and $2.03$ MeV, respectively. Meanwhile, the $\bar{K}^{*}B$  component decreases from approximately $45\%$ to $40\%$ and $37\%$, respectively, while the $\bar{K}^{*}B^{*}$ component increases correspondingly.

Our results also reveal another feature of channel coupling: when a coupled-channel state is dominated by one channel, its binding properties tend to resemble those obtained in the corresponding single-channel calculation. As shown in Table~\uppercase\expandafter{\romannumeral 6}, the molecular states in the $J^P=1^{+}$, $2^{\pm}$, and $3^{\pm}$ channels obtained by including channel coupling are all dominated by the $\bar{K}^{*}B^{*}$ component. This dominance is particularly pronounced in the $J^P=1^{+}$ and $3^{-}$ channels, where the $\bar{K}^{*}B^{*}$ component accounts for nearly $100\%$ of the molecular state. By comparing these results with those obtained from the two single-channel interactions, we can further examine the effects of channel coupling on the binding properties. Taking the $J^P=2^{-}$ channel as an example, the coupled-channel calculation yields a molecular state with a binding energy of $1.81$ MeV at $\alpha=1.1$, with the $\bar{K}^{*}B$ and $\bar{K}^{*}B^{*}$ components accounting for $17\%$ and $83\%$, respectively. In contrast, the pure $\bar{K}^{*}B^{*}$ molecular state has a binding energy of only $0.73$ MeV at the same value of $\alpha$, while no bound state is found in the pure $\bar{K}^{*}B$ channel.  We also find that channel coupling can lead to more deeply bound molecular states. Taking the $J^P=3^{+}$ channel as an example, the binding energies of the coupled $\bar{K}^{*}B/\bar{K}^{*}B^{*}$ molecular state and the pure $\bar{K}^{*}B^{*}$ molecular state at $\alpha=2.0$ are $1.38$ MeV and $0.93$ MeV, respectively.

\begin{figure}
    \centering
    \includegraphics[width=1\linewidth]{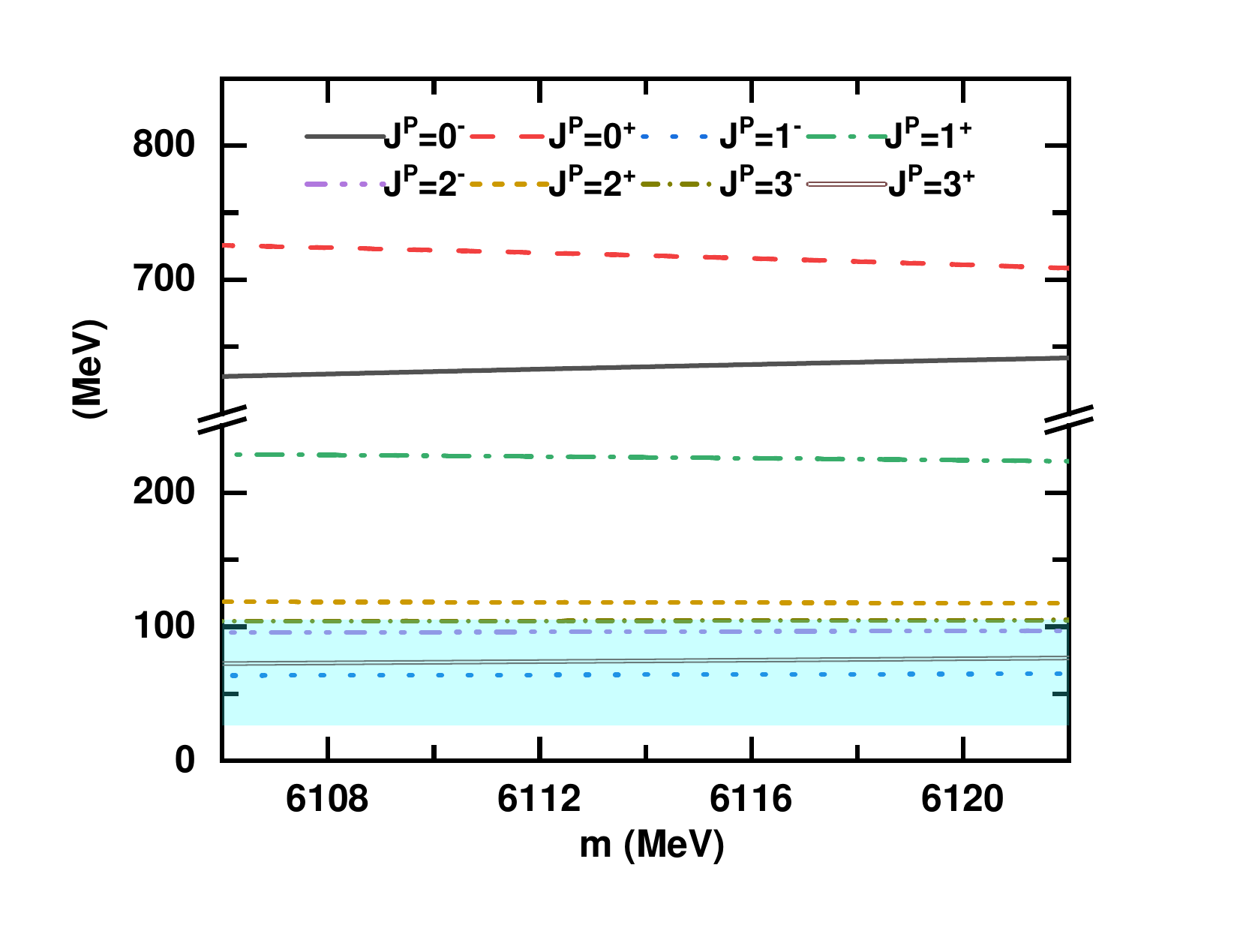}
    \caption{ Decay widths of the $B^{(*)}\bar{K}^{*}$ molecular states as functions of their masses. The cyan band represents the experimental measurement, which are obtained form Ref.~\cite{ParticleDataGroup:2024cfk}.}
    \label{width}
\end{figure}

Based on the mass spectrum obtained in our calculations, the $B_{sJ}(6114)$ state can be interpreted as a $\bar{K}^{*}B^{(*)}$ molecular state; however, its spin-parity quantum numbers cannot yet be determined. Therefore, within the present framework, we calculate its decay widths in the hope of helping identify its spin-parity assignment. These calculations are performed using the coupled-channel results. The wave functions of the $\bar{K}^{*}B$ and $\bar{K}^{*}B^{*}$ components for different spin-parity assignments, as defined in Eq.~\ref{eq24}, are shown in Fig.~\ref{fig:decay_processes}. For clarity, these wave functions are presented in momentum space.  The calculated decay widths as functions of the molecular-state mass are presented in Fig.~\ref{width}, where the molecular-state masses are computed relative to the threshold of the dominant $B^{\*}\bar{K}^{\*}$ channel. As the molecular-state mass approaches that of $B_{sJ}(6114)$, the $B^{\*}\bar{K}^{\*}$ component exceeds $89\%$ in all cases.

As shown in the figure, the predicted total decay width for the $J^P=1^{+}$ assignment is much larger than the experimental value, disfavoring the interpretation of $B_{sJ}(6114)$ as a $\bar{K}^{*}B$--$\bar{K}^{*}B^{*}$ molecular state with $J^P=1^{+}$. This assignment corresponds to an $S$-wave-dominated molecular state. Similarly, the $J^P=0^{+}$ and $J^P=0^{-}$ assignments for the $B^{(*)}\bar{K}^{*}$ molecular configurations can also be excluded, as their predicted decay widths are also much larger than the experimental value. Furthermore, the $J^P=2^{+}$ and $J^P=3^{-}$ assignments for the $\bar{K}^{*}B$--$\bar{K}^{*}B^{*}$ molecular configuration are also disfavored, as their predicted decay widths slightly exceed the upper limit of the experimentally allowed range.  It is worth noting that the $J^P=2^+$ state contains only a small $S$-wave component.

These results suggest that $B_{sJ}(6114)$ is unlikely to be an $S$-wave-dominated $\bar{K}^{*}B$--$\bar{K}^{*}B^{*}$ molecular state and may instead be dominated by higher partial waves. In particular, the $J^P=2^{-}$, $J^P=3^{+}$, and $J^P=1^{-}$ assignments remain viable, as their predicted decay widths fall within the experimentally allowed range. These assignments correspond to molecular states dominated by the $P$-wave, $D$-wave, and $P$-wave contributions, respectively. Notably, the predicted decay width for the $J^P=1^{-}$ assignment is closest to the experimental central value.

\section{\label{sec:level4}summary}
The observation of the bottom-strange meson $B_(6114)$ by the LHCb Collaboration has stimulated considerable theoretical interest in its nature. Although various quark-model assignments, including the $1^3D_1$ state, have been proposed, its spin-parity quantum numbers remain experimentally undetermined, and its possible interpretation as a hadronic molecular state has yet to be systematically investigated.

In this work, we investigate whether $B_{sJ}(6114)$ can be interpreted as a $B^{(*)}\bar{K}^{*}$ molecular state. Using the one-boson-exchange model and the effective Lagrangian approach, we derive the effective potentials arising from pseudoscalar and vector meson exchanges. With monopole form factors introduced at the interaction vertices, we solve the nonrelativistic Schrödinger equation to search for bound states. Using the resulting bound-state wave functions, we then calculate the corresponding strong decay widths into the allowed two-body $B^{(*)}\bar{K}$ channel.

Our numerical results show that the $B^{(*)}\bar{K}^{*}$ interaction can support bound states with various spin-parity assignments, suggesting that $B_{sJ}(6114)$ may be interpreted as a $B^{(*)}\bar{K}^{*}$ molecular state. To constrain its spin-parity quantum numbers, we compare the predicted decay widths with the experimental measurements. We find that the $J^P=1^-$, $2^-$, and $3^+$ assignments are favored over other possibilities. These assignments suggest that $B_{sJ}(6114)$ could be a $B^{(*)}\bar{K}^{*}$ molecular state dominated by higher partial waves. Among these possibilities, the $J^P=1^-$ assignment provides the best agreement with the experimental data, corresponding to a molecular state dominated by the $P$-wave component.

\section{\label{sec:level1}acknowledgments}
This work was supported by the Fundamental
Re search Funds for Central Universities under Grant $No.
2682026TPY011$.


\newpage
\begin{thebibliography}{99}
\bibitem{ParticleDataGroup:2024cfk}
S.~Navas \textit{et al.} [Particle Data Group],
Phys. Rev. D \textbf{110} (2024), 030001.


\bibitem{Godfrey:1985xj}
S.~Godfrey and N.~Isgur,
Phys. Rev. D \textbf{32}, 189-231 (1985).



\bibitem{Guo:2006fu}
F.~K.~Guo, P.~N.~Shen, H.~C.~Chiang, R.~G.~Ping and B.~S.~Zou,
Phys. Lett. B \textbf{641} (2006), 278-285.



\bibitem{Guo:2006rp}
F.~K.~Guo, P.~N.~Shen and H.~C.~Chiang,
Phys. Lett. B \textbf{647} (2007), 133-139.



\bibitem{Altenbuchinger:2013vwa}
M.~Altenbuchinger, L.~S.~Geng and W.~Weise,
Phys. Rev. D \textbf{89} (2014), 014026.


\bibitem{Meng:2022ozq}
L.~Meng, B.~Wang, G.~J.~Wang and S.~L.~Zhu,
Phys. Rept. \textbf{1019}, 1-149 (2023).



\bibitem{BaBar:2003oey}
B.~Aubert \textit{et al.} [BaBar],
Phys. Rev. Lett. \textbf{90} (2003), 242001.


\bibitem{CLEO:2003ggt}
D.~Besson \textit{et al.} [CLEO],
Phys. Rev. D \textbf{68} (2003), 032002
[erratum: Phys. Rev. D \textbf{75} (2007), 119908].



\bibitem{Belle:2003kup}
Y.~Mikami \textit{et al.} [Belle],
Phys. Rev. Lett. \textbf{92} (2004), 012002.



\bibitem{Belle:2003guh}
P.~Krokovny \textit{et al.} [Belle],
Phys. Rev. Lett. \textbf{91} (2003), 262002.



\bibitem{BaBar:2004yux}
B.~Aubert \textit{et al.} [BaBar],
Phys. Rev. Lett. \textbf{93} (2004), 181801.



\bibitem{Xie:2010zza}
Z.~X.~Xie, G.~Q.~Feng and X.~H.~Guo,
Phys. Rev. D \textbf{81} (2010), 036014.

\bibitem{Gamermann:2006nm}
D.~Gamermann, E.~Oset, D.~Strottman and M.~J.~Vicente Vacas,
Phys. Rev. D \textbf{76} (2007), 074016.


\bibitem{Zhu:2019vnr}
H.~Zhu and Y.~Huang,
Phys. Rev. D \textbf{100} (2019), 054031.


\bibitem{Mohler:2013rwa}
D.~Mohler, C.~B.~Lang, L.~Leskovec, S.~Prelovsek and R.~M.~Woloshyn,
Phys. Rev. Lett. \textbf{111} (2013), 222001.

\bibitem{Isgur:1991wq}
N.~Isgur and M.~B.~Wise,
Phys. Rev. Lett. \textbf{66}, 1130-1133 (1991).

\bibitem{Wang:2019ehs}
Z.~Y.~Wang, J.~J.~Qi, Q.~X.~Yu and X.~H.~Guo,
Phys. Rev. D \textbf{100} (2019), 096009.


\bibitem{Cheng:2014bca}
H.~Y.~Cheng and F.~S.~Yu,
Phys. Rev. D \textbf{89} (2014), 114017.



\bibitem{Colangelo:2012xi}
P.~Colangelo, F.~De Fazio, F.~Giannuzzi and S.~Nicotri,
Phys. Rev. D \textbf{86} (2012), 054024.



\bibitem{Kolomeitsev:2003ac}
E.~E.~Kolomeitsev and M.~F.~M.~Lutz,
Phys. Lett. B \textbf{582} (2004), 39-48.


\bibitem{Yang:2022vdb}
Z.~Yang, G.~J.~Wang, J.~J.~Wu, M.~Oka and S.~L.~Zhu,
JHEP \textbf{01} (2023), 058.


\bibitem{Lang:2015hza}
C.~B.~Lang, D.~Mohler, S.~Prelovsek and R.~M.~Woloshyn,
Phys. Lett. B \textbf{750} (2015), 17-21.


\bibitem{Albaladejo:2016ztm}
M.~Albaladejo, P.~Fernandez-Soler, J.~Nieves and P.~G.~Ortega,
Eur. Phys. J. C \textbf{77} (2017), 170.



\bibitem{Vishwakarma:2022sly}
K.~K.~Vishwakarma and A.~Upadhyay,
Eur. Phys. J. A \textbf{58} (2022), 38.


\bibitem{D0:2016mwd}
V.~M.~Abazov \textit{et al.} [D0],
Phys. Rev. Lett. \textbf{117},, 022003 (2016).



\bibitem{ATLAS:2018udc}
M.~Aaboud \textit{et al.} [ATLAS],
Phys. Rev. Lett. \textbf{120}, 202007 (2018).


\bibitem{CMS:2017hfy}
A.~M.~Sirunyan \textit{et al.} [CMS],
Phys. Rev. Lett. \textbf{120} (2018), 202005.



\bibitem{LHCb:2015aaf}
R.~Aaij \textit{et al.} [LHCb],
JHEP \textbf{04}, 024 (2015).


\bibitem{LHCb:2016dxl}
R.~Aaij \textit{et al.} [LHCb],
Phys. Rev. Lett. \textbf{117}, 152003 (2016).


\bibitem{CDF:2017dwr}
T.~Aaltonen \textit{et al.} [CDF],
Phys. Rev. Lett. \textbf{120} (2018), 202006.
















































\bibitem{LHCb:2020pet}
R.~Aaij \textit{et al.} [LHCb],
Eur. Phys. J. C \textbf{81} (2021) no.7, 601
doi:10.1140/epjc/s10052-021-09305-3
[arXiv:2010.15931 [hep-ex]].

\bibitem{Patel:2022hhl}
V.~Patel, R.~Chaturvedi and A.~K.~Rai,
[arXiv:2201.01120 [hep-ph]].

\bibitem{Chen:2022fye}
B.~Chen, S.~Q.~Luo, K.~W.~Wei and X.~Liu,
Phys. Rev. D \textbf{105} (2022) no.7, 074014
doi:10.1103/PhysRevD.105.074014
[arXiv:2201.05728 [hep-ph]].

\bibitem{Gandhi:2022nnk}
K.~Gandhi and A.~K.~Rai,
Eur. Phys. J. C \textbf{82} (2022) no.9, 777
doi:10.1140/epjc/s10052-022-10719-w
[arXiv:2208.08791 [hep-ph]].

\bibitem{Garg:2023hvp}
R.~Garg, P.~Gupta and A.~Upadhyay,
PTEP \textbf{2023} (2023) no.7, 073B03
doi:10.1093/ptep/ptad080
[arXiv:2305.02557 [hep-ph]].

\bibitem{Jakhad:2025ejc}
P.~Jakhad and A.~K.~Rai,
Phys. Rev. D \textbf{112} (2025) no.7, 074011
doi:10.1103/m2xt-pkxm
[arXiv:2509.14699 [hep-ph]].

\bibitem{Sanchez-Illana:2026cyv}
J.~S{\'a}nchez-Illana, R.~Molina and P.~P.~Shi,
[arXiv:2603.28649 [hep-ph]].

\bibitem{Hiyama:2003cu}
E.~Hiyama, Y.~Kino and M.~Kamimura,
Prog. Part. Nucl. Phys. \textbf{51} (2003), 223-307.

\bibitem{Chen:2021tip}
R.~Chen,
Eur. Phys. J. C \textbf{81} (2021), 122.


\bibitem{wise1992}
M.~B.~Wise,
Phys. Rev. D \textbf{45}, R2188 (1992).

\bibitem{yan1992}
T.-M.~Yan, H.-Y.~Cheng, C.-Y.~Cheung, G.-L.~Lin, Y.~C.~Lin, and H.-L.~Yu,
Phys. Rev. D \textbf{46}, 1148 (1992); \textbf{55}, 5851(E) (1997).

\bibitem{grinstein1992}
B.~Grinstein, E.~E.~Jenkins, A.~V.~Manohar, M.~J.~Savage, and M.~B.~Wise,
Nucl. Phys. B \textbf{380}, 369 (1992).

\bibitem{casalbuoni1997}
R.~Casalbuoni, A.~Deandrea, N.~Di~Bartolomeo, R.~Gatto, F.~Feruglio, and G.~Nardulli,
Phys. Rep. \textbf{281}, 145 (1997).

\bibitem{li2012}
N.~Li and S.-L.~Zhu,
Phys. Rev. D \textbf{86}, 074022 (2012).

\bibitem{li2013}
N.~Li, Z.-F.~Sun, X.~Liu, and S.-L.~Zhu,
Phys. Rev. D \textbf{88}, 114008 (2013).

\bibitem{molina2010}
R.~Molina, T.~Branz, and E.~Oset,
Phys. Rev. D \textbf{82}, 014010 (2010).

\bibitem{isola2003}
C.~Isola, M.~Ladisa, G.~Nardulli, and P.~Santorelli,
Phys. Rev. D \textbf{68}, 114001 (2003).

\bibitem{bando1988}
M.~Bando, T.~Kugo, and K.~Yamawaki,
Phys. Rep. \textbf{164}, 217 (1988).

\bibitem{Xie:2015zga}
J.~J.~Xie, Y.~B.~Dong and X.~Cao,
Phys. Rev. D \textbf{92} (2015), 034029.

\bibitem{He:2011jp}
J.~He, Z.~Ouyang, X.~Liu and X.~Q.~Li,
Phys. Rev. D \textbf{84} (2011), 114010.


\bibitem{Liu:2006df}
X.~Liu, B.~Zhang and S.~L.~Zhu,
Phys. Lett. B \textbf{645} (2007), 185-188.



\bibitem{Colangelo:2002mj}
P.~Colangelo, F.~De Fazio and T.~N.~Pham,
Phys. Lett. B \textbf{542} (2002), 71-79.



\bibitem{Meng:2007cx}
C.~Meng and K.~T.~Chao,
Phys. Rev. D \textbf{75} (2007), 114002.


\bibitem{Xu:2015qqa}
H.~Xu, J.~J.~Xie and X.~Liu,
Eur. Phys. J. C \textbf{76} (2016), 192.


\bibitem{Chen:2012nva}
D.~Y.~Chen, X.~Liu and T.~Matsuki,
Phys. Rev. D \textbf{87} (2013), 054006.


\end{thebibliography}
\end{document}